%Paper: hep-th/9204090
%From: Malcolm Perry <M.J.Perry@damtp.cambridge.ac.uk>
%Date: Tue, 28 Apr 92 16:41:45 BST

\input harvmac
\Title{hep-th/9204090}{The Physics of 2-d Stringy Spacetimes}
\centerline{G.W. Gibbons\footnote{$^\dagger$}
{(gwg1@uk.ac.cam.phx)}}
\centerline{DAMTP,}\vskip-0.06in
\centerline{University of Cambridge,}\vskip-0.06in
\centerline{Silver Street,}\vskip-0.06in
\centerline{Cambridge CB3 9EW,}\vskip-0.06in
\centerline{England.}
\bigskip
\centerline{Malcolm J. Perry\footnote{$^\ddagger$}
{(malcolm@amtp.cam.ac.uk), on leave from DAMTP, University of Cambridge.}}
\centerline{Department of Mathematics,}\vskip-0.06in
\centerline{Massachusetts Institute of Technology,}\vskip-0.06in
\centerline{Cambridge, MA 02139,}\vskip-0.06in
\centerline{USA.}
\vskip .3in
We examine the two-dimensional spacetimes that emerge from string theory.
We find all of the solutions with no tachyons, and show that the only
non-trivial solution is the black hole spacetime. We examine the role of
duality in this picture. We then explore the thermodynamics of these
solutions which is complicated by the fact that only in two spacetime
dimensions is it impossible to redefine the dilaton field in terms of a
canonical scalar field. Finally, we extend our analysis to the heterotic
string, and briefly comment on exact, as opposed to perturbative,
solutions.

\Date{3/92}

\def\bxy{{\cal\char'064}\,}
\newsec{Introduction}

The two dimensional black hole that emerges from the ${SU(1,1)/ U(1)}$
\lq\lq coset" model \ref\GKO{P. Goddard, A. Kent and D.I. Olive, Comm.
Math. Phys. {\bf 103}, 105, 1986.}
is an interesting testing ground for many ideas about
the physics of black holes. In general relativity, black holes behave as
perfect absorbers. If one approximates quantum gravity by treating
quantum fields as propagating on a classical background geometry,
the semiclassical approximation,  one then finds that the black hole
behaves as a perfect black body  with a temperature $T_H$, the Hawking
temperature, \ref\Hawk{S.W. Hawking, Comm.Math. Phys.
{\bf 43}, 199, 1975.}.
For an uncharged, non-rotating black hole in four dimensional
spacetime, whose geometry is completely determined by the mass of the
hole $M$
\eqn\THawk{ T_H = {1 \over 8\pi M}}
Hawking's result has profound consequences for any quantum theory of
gravity. Suppose that it is true for all $M$, then a black hole is
unstable as its specific heat is negative. Thus the black hole will
radiate, lose mass and consequently increase its temperature, resulting in
the catastrophic disappearance of the black hole. It would then
appear that there is no unitary time evolution operator for quantum
gravity, \ref\breakdown{S.W. Hawking, Phys. Rev. {\bf D14}, 2460, 1976.}.
The incoming state which gives rise to the black hole can be chosen to be
a pure state with zero entropy. The outgoing state is a mixed thermal
state, with large entropy. Since quantum mechanical evolution preserves
entropy, it is often  argued that quantum gravity must somehow transcend
the usual laws of quantum mechanics.

Various suggestions have been made for side-stepping this problem.
One is to suppose that \THawk~ only holds for large $M$, and in fact
a small black hole will persist as a repository for all the quantum
mechanical information that would have been lost had the black hole
completely disappeared. At first sight, it seems to be  unlikely
as  small black holes must then contain arbitrarily large amounts of
information. The initial black hole could have had a very
large mass, being formed from a large quantity of matter and hence
have a very large entropy.
One can be a little more precise in seeing what is wrong
with this suggestion.
In order for \THawk~ to be
consistent with the first law of thermodynamics, the entropy $S$
\Hawk, \ref\BCH{J. Bardeen, B. Carter and S.W. Hawking, Comm. Math. Phys.
{\bf 31}, 161, 1973.},
 \ref\Bek{J. Bekenstein, Phys. Rev. {\bf D7}, 2333, 1973.}
 of the
hole must be given by
\eqn\SBek { S = 4\pi M^2.}
If we use the Boltzmann interpretation of entropy, then the density
of states function is the exponential of $S$, and thus small mass black
holes have a small density of states and hence can contain only a little
information.

Another suggestion is that the reason one finds non-unitary time
evolution is that our description of the initial state has had all
the quantum mechanical information removed from it by virtue of thinking
about it as a classical field configuration.  It would seem to be
inevitable that such problems appear in the semi-classical
approximation. It might then appear that the radiation is approximately
thermal in the same way that radiation from the sun is thermal.
Such radiation is described by a mixed state because we cannot keep track
of all the quantum mechanical information involved in a very complicated
system. In order to see whether this is the case, we would need a
quantum theory of gravitation that makes sense at a fundamental level.
String theory offers one such candidate, although
only at the level of perturbation theory. However, we have so far
ignored problems associated with the inherent indescribability
of the quantum state of matter inside the black hole. For example,
a device recording information about what is falling into the hole
cannot communicate its observations to any observer exterior to the hole.
Perhaps a resolution of our difficulties is that a global construct like
the event horizon is inherently an un-quantum mechanical concept.

Yet another possible resolution is to blame the loss of quantum coherence
on the spacetime singularity that  is inevitably  found interior to the
black hole, at least in classical general relativity.
Experience with other areas of physics indicates that
singularities are fictions of classical physics, and quantum
mechanically the  problems  associated with them will disappear.
There is a  concrete suggestion
that this is indeed the case  for two-dimensional black holes,
\ref\Wittenbh{E. Witten, Phys. Rev. {\bf D44 }, 314, 1991.}.
An understanding of what becomes of singularities in a
quantum theory of gravity will then give us the key to the resolution of
our difficulties.

A quite different
way round our difficulties is to adopt a suggestion of Dyson
\ref\dyson{F.J. Dyson, Institute for Advanced Study preprint,
1976.}, namely that
the black hole precipitates the formation of a baby universe that contains
the quantum mechanical information associated with the gravitational
collapse. The \lq\lq bag of gold" type of construction would replace the
singularity in the interior of the black hole and allow the Hawking
process to proceed producing entropy exterior to the black hole.
Dyson's picture allows for  unitary time evolution in quantum gravity as
the  hidden information would be preserved in the baby universe.
The price paid  is that the time reverse of such a process is quite
possible and would be interpreted as a white hole. Such things do not
appear to occur in our universe.

Above, we have sketched some of the possible ideas
about one of the central issues in quantum gravity today,
 however our list
is clearly far from describing all the possiblities.
Whatever the resolution to these problems, it is clear that we do not have a
complete consistent quantum theory of gravitation which is able to make
reliable predictions. However, it seems that elementary thermodynamic
considerations can shed some  light on the fundamental difficulties of
quantum gravity.

String theory is one approach to quantum gravity. In its present
form, strings move in  certain fixed classical backgrounds. These
backgrounds are determined by asking that the conformal invariance of
the string is not broken by the background in which it moves. to requiring
Such an approach has much in common
with the semi-classical analysis because the classical spacetime
background and the fields propagating on it are regarded as string
condensates, which are inherently classical in the only available
formulation of string theory, namely a first quantized string theory.

In some of the scenarios we described, or minor variants of them,
one might expect black hole solitons
to emerge as field configurations which have zero Hawking temperature,
and a small intrinsic entropy. In general relativity, extreme
Reissner-Nordstrom black holes are examples of such configurations
\ref\sols{G.W. Gibbons, {\it Unified Theories of Elementary
Particles}, eds P. Breitenlohner and H.P. D\"urr, Springer Verlag,
Heidelburg, 1982.}. In fact, it might be rather satisfying to find solitons
with zero entropy as these would correspond to pure quantum mechanical
states and hence be some kind of fundamental excitation associated with the
gravitational field.

In the remainder of this paper, we discuss stringy black holes in
an attmept to use elementary methods to shed  light on these
perplexing issues. In section two, we describe the black hole spacetimes
that emerge from the theory of closed bosonic strings. The system of
equations
that emerge describing the low-energy stringy backgrounds turns out to be
exactly integrable in the absence of any tachyon fields. There are seven
local solutions to the equations, the metrics of which turn out to be
either flat spacetime, different coordinate patches on the black hole
spacetime or derivable from them. In
section three, we explore the thermodynamic interpretation of the
spacetimes found in section three. In section four, we extend our
discussion to the case of the stringy backgrounds that emerge from
heterotic string theory.  Rather disaapointingly, we discover
that there are no stable fundamental objects to be
found in this theory, despite the superficial similarities to four
dimensional Einstein-Maxwell theory. Finally in section five, we very
briefly outline the exact string backgrounds which arise from the coset
space construction. The paper ends with some conclusions and speculations
on the role of black holes in string theory and the consequences for a
theory of quantum gravity.

\newsec{Two dimensional stringy backgrounds}

We assume that the closed bosonic string is moving in a two-dimensional
background spacetime. If we look at the physical state conditions, we
find that the only conventional physical mode  is the spacetime
\lq\lq tachyon" $T$, which because of the two-dimensionality is in fact
massless. However there are other string states, the so-called topological
states \ref\polaykov{A. M. Polyakov, Mod. Phys. Lett. {\bf
A6}, 635, 1991.}, which come from the higher modes of the string whose
longitudinal components survive as background fields, even though
they vanish completely as dynamical degrees of freedom.
These modes are represented the background gravitational field
specified by the metric $g_{ab}$, and the dilaton field $\Phi$.

The string $\beta$-functions then, to lowest order
\ref\betafuncs{C. Callan, D. Friedan, E. Martinec and M.J. Perry,
Nucl. Phys. {\bf B262}, 593, 1985.}, are
\eqn\betag{ R_{ab} = \nabla_a~\nabla_b\Phi + \nabla_aT\nabla_bT}
\eqn\betaphi{ R - (\nabla \Phi)^2 - 2\bxy\Phi - (\nabla T)^2 - V(T)
= c^\ast,}
\eqn\betaT{\bxy T + \nabla_a \Phi~ \nabla^a T = {1\over 2}V'(T),}
where, $V(T)$ is the \lq\lq tachyon" potential and $c^\ast$ is related to
effective central charge of the world sheet fields. It should be
emphasized that to higher orders in $\alpha^\prime$, there will
be higher order terms appearing which we will ignore in what follows.
The bosonic
string can be coupled to any conformal field theory with central charge
$c$,
in which case the effective central charge is
\eqn\ceffect{c^\ast = {c + d -26 \over 3}}
for a closed bosonic string in $d$-dimensions.
The factor of $26$ comes from
the reparametrization ghosts, and $d$ comes from the $d$ string
coordinates which behave like free bosons, at least in flat spacetime.
 It should be noted that for the
minimal bosonic
string theory, not coupled to any extra worldsheet fields,
that $c^\ast < 0$.
The tachyon potential $V(T)$ takes the form
\eqn\tachpot{  V(T) = V_0 -2T^2 + V_3T^3 + O(T^4) }
All of these equations can be derived from a target space action
\eqn\tgtspaction{ \int_{\cal M} d^2x g^{1\over2}e^\phi\biggl(c^\ast - R
-(\del\phi)^2+(\del T)^2 + V(T)\biggr) - 2\int_{\del\cal M} e^\phi K
d\Sigma}
Extremizing this action with respect to arbitrary variations of the fields
within the region of spacetime ${\cal M}$, but with the
value of the dilaton and metric induced on the
boundary, $\del\cal M$, fixed will reproduce the $\beta$-function equations.
In the boundary term, $K$ is the trace of the second fundamental form
of the boundary. Omission of this boundary term will fail to
yield a well-posed
variational principle that involves only the fields and their
first derivatives, ~\ref\York{J.A. York, Phys. Rev. Lett. {\bf 28}, 1082,
1972.}, \ref\GibbHaw{G.W. Gibbons and S.W. Hawking, Phys. Rev.
{\bf D15}, 2752, 1977.}.

We may as well set $V_0=0$   since in string perturbation theory it vanishes.
If it does not vanish outside perturbation theory, it can be absorbed into
a the definition of $c^\ast$. $V_3$ has been calculated by
Samuel \ref\kost{ S. Samuel, Phys. Lett. {\bf 181B}, 255, 1985.}
but  we ignore it as it is higher order in
$\alpha^\prime$.

We now consider the solutions of the equations \betag,~\betaphi~ and
\betaT~ in two spacetime dimensions. We treat the case where the
tachyon condensate is put to zero, $T=0$. \betag~ then implies the
existence of an isometry. In $d=2$, the Ricci tensor obeys the Gauss
identity $R_{ab}={1\over 2}Rg_{ab}$, so \betag~
can be rewritten as
\eqn\confdil{\nabla_a\nabla_b\phi - {1\over 2}g_{ab} \bxy\phi =0}
Thus $\nabla_a\phi$ is a conformal Killing vector. A consequence of this
also special to two dimensions, is that the vector dual to this,
\eqn\dualphi{k_a = \epsilon_a{}^b \nabla_b\phi}
where $\epsilon_{ab}$ is the two-dimensional alternating tensor,
obeys Killing's equation
\eqn\Kill{\nabla_ak_b+\nabla_bk_a=0}

 Thus all solutions
to these equations must have
at least one Killing vector. A further consequence of this is that
\eqn\staticphi{k^a \nabla_a \phi = 0 }
This means that the level sets of $\phi$ are the orbits of the Killing
vector.

We can now construct all local solutions to these equations.
Suppose that we
start with the metric written in  form
\eqn\genmetric{ ds^2 = -f^2(r)dt^2 + dr^2, }
so that $k^a{\partial\over\partial x^a}={\partial\over\partial t}$.
We are guaranteed to be able to write the metric in this form locally
by virtue of the
existence of the Killing vector $k_a$. Furthermore, it follows that
$\phi$ must be a function of $r$ only be virtue of \staticphi.
An interesting property of solutions of the string equations
of this form is that of duality, \ref\duality{
T.H.Buscher, Phys. Lett. {\bf 194B}, 59, 1987;
Phys. Lett. {\bf 201B}, 466, 1988.}.
 This is a map that transforms a solution
of the beta-function equations,
\betag~ and \betaphi~ with zero tachyon field, into a related solution
of the same equations. In terms of the variables in \genmetric,
duality effects the following transformations
\eqn\duality{ f \rightarrow {1\over f}, \ \ \ \ \ \phi \rightarrow \phi
+2\ln f.}
It is believed that this duality transformation is a feature of the
exact theory, and not just an artefact of the low-order approximation
to the beta-function equations.

The simplest solution of the field equations is flat spacetime. This
corresponds to the choice
\eqn\flat{ f(r) = 1, \ \ \ \ \ \phi=\phi_0 + 2\lambda (r-r_0)}
Here, both $\phi_0$ and $r_0$ are constants of integration. $\lambda$
is a constant that is determined by the central charge, and is given by
\eqn\cent{\lambda = \sqrt{-c^\ast\over 4}}
This shows us a further interesting relationship, namely that for
$k^a$ to be timelike, since it is the Hodge dual of $\nabla_a\phi$, $\phi$
can only be a function of $r$, and then since $\phi$ is a real field,
$c^\ast$ must be negative. One might also think that it is possible to find
solutions with $c^\ast>0$, but this would require us to violate the no-ghost
theorem.

We choose to fix the redundant parameter $r_0$ by requiring
\eqn\defrzero{r_0 = {1\over \lambda}\ln 2.} It should be noted that
this spacetime is transformed into itself under the duality transformation.

A second  solution of these equations is the metric exterior to a black
hole.
This was first discovered in
\ref\Rocek{ M. Rocek, K. Schoutens and A. Sevrin, Phys. Lett. {\bf 256B,}
303, 1991.},
\ref\Wadia{G. Mandal, A.M. Sengupta  and S.R. Wadia, Mod. Phys. Lett. {\bf
A6}, 1685, 1991.}.  In our coordinates, we find \eqn\bh{ f(r) =
\tanh\lambda(r-r_0),\ \ \ \ \ \phi=\phi_0 + 2\ln\cosh \lambda(r-r_0)}
This spacetime is asymptotic to the flat space solution of \flat.
The value of $\lambda$ is determined by the central charge exactly as
for the flat space case. The solution has a horizon bifurcation point
at the location of the coordinate singularity at $r=r_0$. In what follows,
we will choose $r_0=0$. The global structure of this spacetime can be made
more transparaent by transforming to Kruskal type null coordinates
$u$ and $v$, when the metric becomes
\eqn\kruskal{ ds^2 = -{1\over\lambda^2}(1-uv)^{-1} dudv, \ \ \ \ \ \phi=\phi_0
+\ln(1-uv).}
This spacetime is thus the two-dimensional analog of the Schwarzschild
solution, with the Kruskal coordinates being given in terms of
$r$ and $t$ by \eqn\krusU{u= -e^{-\lambda t}\sinh\lambda r,}
\eqn\krusV{v = e^{\lambda t}\sinh\lambda r.} The horizons are
the null surfaces where $u$ or $v$ vanish. The line $uv=1$
is the spacetime singularity. If we were talking about general relativity,
this singularity would be the boundary of spacetime. This interpretation
arises because we are compelled to assign boundary conditions at the
singularity
in order to define physics in its Cauchy development.
However, the physical meaning here is somewhat different, because
although the spacetime is singular, the associated conformal field theory
is claimed to be well-behaved, \Wittenbh.
 The singularity does not therefore form
a barrier to passing into new regions of spacetime which contain naked
singularities. The maximal extension of the spacetime thus contains six
distinct regions, two asymptotically flat regions exterior to the horizons,
($u<0~,v>0$ which is our original region, and $u>0~,v<0$)
two  regions which are interior to the horizons ($u,v>0; uv<1$ and
$u,v<0; uv<1$) and bounded by the
singularities, and finally  two regions which are asymptotically flat
and contain a naked singularity, ($u,v>0; uv>1$ and  $u,v<0; uv>1$).
 These last regions have no analog in the
Kruskal manifold found in four dimensional general relativity.

The black hole spacetime can be transformed by the duality transformations
into a different asymptotically flat spacetime \ref\giveon{
A. Giveon, Mod. Phys. Lett. {\bf A6,} 2843, 1991.}, sometimes called the
trumpet spacetime,
\eqn\trumpet{ f(r)=\coth\lambda r, \ \ \ \ \phi=\phi_0 +
2\ln\sinh\lambda r.} This metric is asymptotically flat and  contains a
naked singularity at $r=0$. In fact, it is isometric to the new regions of
the Kruskal manifold as can be seen by making the coordinate
transformations \eqn\trumpu{u=\pm\exp(-\lambda t)\cosh\lambda r,}
\eqn\trumpv{v=\pm\exp(\lambda t)\cosh\lambda r,}
which take \trumpet~ into \kruskal.

Another pair of metrics that are dual to each other are given
by
\eqn\elliphole{ f(r)=\tan\lambda r, \ \ \ \ \phi=\phi_0 - 2\ln\cos\lambda r,}
and
\eqn\elliptrump{ f(r) = \cot\lambda r, \ \ \ \ \phi=\phi_0 - 2\ln\sin
\lambda r.}
These are in fact also  diffeomorphic to each other as can be seen making
the coordinate transformation
\eqn\elliptr{ r \rightarrow r + {\pi\over 2\lambda}}
and renormalizing $\phi_0$ appropriately. Both of these solutions can
also be found by making a complex coordinate transformation on the black
hole or naked singularity spacetimes, \bh~ and \trumpet. This pair of
spacetimes are both singular where $f(r)=0$, and are not asymptotically
flat. However, they are the regions of the Kruskal manifold corresponding
to the spacetime interior to the black hole horizon. In \elliphole~
the  surface $r=0$ is the singularity of the Kruskal manifold,
and $r={\pi\over2\lambda}$ is the horizon. For \elliptrump~
the the singularity is at $r={\pi\over2\lambda}$ and the horizon is
at $r=0$. To exhibit the relation explicitly,
we can take the transformations
\eqn\kruskbhu{ u = \pm\exp(-\lambda t)\sin\lambda r,}
\eqn\kruskbhv{ v = \pm\exp(\lambda t)\sin\lambda r,}
which takes the Kruskal metric into \elliphole
and
\eqn\krusktrumpu{ u = \pm\exp(-\lambda t)\cos\lambda r,}
\eqn\krusktrumpv{ v = \pm\exp(\lambda t)\cos\lambda r,}
which takes the Kruskal metric into \elliptrump.

There is a final pair of local solutions to the equations \genmetric.
One can choose the metric on flat space to be given by the Rindler form
instead of
\flat~
\eqn\rindler{ f(r) =  r, \ \ \ \ \phi=\phi_0 - 2\ln r.}
One could also have derived this by taking the black hole solution \bh~
and taking the $\lambda$ goes to zero limit whilst simultaneously rescaling
the $t$-coordinate. Clearly, there is a spacetime that is dual to it,
namely
\eqn\horn { f(r) = {1\over r}, \ \ \ \ \phi=\phi_0 + 2 \ln r.}
This could have been found by a similar limiting procedure applied now to
the trumpet spacetime instead of the black hole spacetime. This spacetime
contains a naked singularity at $r=0$.

All of these metrics have Euclidean continuations. We will only concentrate
on two examples. Firstly Euclidean flat space
is
\eqn\eufs{ds^2 = d\tau^2 + dr^2 \ \ \ \ \phi=\phi_0 +2\lambda r -
\lambda\ln 2}
This can be given the topology of a cylinder,
$S^1 \times R^1$
where the Euclidean time is
periodic with arbitrary period, and $r$ is a coordinate along the cylinder.
This is to be compared and contrasted to the Euclidean black hole
solution, or cigar as it is more popularly known, given by
\eqn\eubh{ ds^2 = \tanh^2\lambda r~ d\tau^2 + dr^2.}
This metric has a conical singularity at $r=0$ unless     we identify
the Euclidean time coordinate $\tau$ with period $ 2\pi\over\lambda$.
The topology of this spacetime is that of a hemisphere.
The local temperature is given by the inverse proper periodicity
of the Euclidean time coordinate. Thus, as
  $r\rightarrow \infty$, the temperature $T_c$
is given by
\eqn\tc{T_c={\lambda\over2\pi}}
In other words, the temperature of the spacetime is determined by the central
charge of the conformal field theory associated with the string theory.

\newsec{Black Hole Thermodynamics}

One way to understand the thermodynamics of gravitational fields is
via the Euclidean treatment of quantum gravity. The basic idea is to
evaluate the partition function as the path integral   over the space
of all field configurations that are periodic in imaginary (that is
to say Euclidean) time. The entropy of black holes in general
relativity is found to originate from the classical contributions to
such a calculation. A similar analysis can be performed for
two-dimensional stringy black holes, and that is the subject
of this section. However we must be careful because as well as
the  gravitational field there is a long-range scalar field, the dilaton,
$\phi$. The dilaton field is associated with both black hole hair, $\phi_0$
and with the background dilaton charge required to cancel the central
charge of the conformal field theory, and thus give rise to
a conformally invariant string theory. The usual way to deal with
a problem of this sort is   to make a conformal transformation
on the spacetime metric so that there is no dilaton field in front
of the $R$ term in the action. The thermodynamics can then be examined
in the usual way
\ref\GibbMaeda{G.W. Gibbons and K-I. Maeda, Nucl. Phys. {\bf B298}, 741,
1988.},
\ref\GibbBreit{G.W. Gibbons and  Breitenlohner, Comm. Math. Phys.
{\bf 120}, 295, 1988. }.
However, in two dimensions this cannot be done, and this
 gives rise to some
complications of the thermodynamics.

To find the classical contribution to the partition function $Z$,
we note that is given by
\eqn\part{ Z=\exp(-I) = \exp(-\beta F)}
where $I$ is the action evaluated for the Euclidean version of
the classical gravitational field in question, $\beta$ is the inverse
temperature, and $F$ is the Helmholtz free energy. If we evaluate the
action for a solution of \betag, \betaphi~ and \betaT~ with $T=0$, we
discover that it  is purely a boundary term,
\eqn\onshellaction{I = -2\int_{\partial {\cal M}} e^\phi(K + n^a\nabla_a
\phi) d\Sigma}
where $n^a$ is the unit outward normal to the boundary  which has $K$
as its second fundamental form and $d\Sigma$ is the volume element on
it.

First we consider the case of flat spacetime. The metric is given by
Euclidean flat space \eufs~, where we identify
the time coordinate $\tau$ with period $\beta$. We will be concerned with
making physical observations at the boundary of the spacetime, in this case
the wall of a box, which is located at a fixed value of the radial
coordinate. However, the value of this coordinate is not
a measureable quantity. The best that we can do is measure
the value of the dilaton field there, which we will call $\phi_W$.
Similarly we can measure the temperature there, $T_W$ which is the
proper periodicity of the Euclidean space at coordinate $r$.
In the flat space example, we find that $T_W = \beta^{-1}$.
Thus $\phi_W$ and $T_W$ are our basic observables, and the
partition function $Z=Z(\phi_W,T_W)$.
As mentioned earlier, we only wish to ask about measurements that
can be made at this wall and address issues about what is happening
inside the box ($r$ decreasing). This means that we will ignore
the second boundary of flat space, that is the boundary at the
component of spacelike
infinity where $r \rightarrow -\infty$.

We now calculate the Euclidean action for flat space from \onshellaction~
and we find
\eqn\fsact{ I = -4{\lambda\over T_W} e^{\phi_W}.}
Flat space has a non-vanishing dilaton charge $D$. Consider the
dilaton current $j_a$
\eqn\dilcurr{ j_a = \epsilon_a{}^b\nabla_be^\phi.}
This current is conserved
\eqn\jcons{ \nabla_a j^a = 0,}
thus there is  an associated conserved charge defined by
\eqn\Dcharge{D = \int_\Sigma j_a d\Sigma^a,}
where $\Sigma$ is a spacelike hypersurface bounded by the wall of the box
and $r=-\infty$. Evaluating the dilaton charge of flat space, we discover
that the dilaton charge inside the box is
\eqn\Dfs{ D=e^{\phi_W},}

Associated with this dilaton charge there will be a chemical potential
$\psi$, the dilaton potential. We can now read off from the action
the value of the Helmholtz free energy F contained in the box in terms of
the coordinates as \eqn\Ffsss{ F = -\lambda e^{\phi_0 + 2\lambda r},}
or in terms of observables
\eqn\Ffss{ F = -4\lambda e^{\phi_W}.}
We should note that $\lambda$ is a fixed constant determined by the
central charge of the string fields.
To find the other thermodynamic potentials, we recall that the first
law of thermodynamics in effect serves to define the chemical potential
and the entropy.
It takes the form here of
\eqn\firstlaw{ dF = -SdT_W - \psi dD}
so $F$ must be expressed in terms of $T_W$ and $D$. Hence
\eqn\Ffs{ F = -\lambda D}
and so the dilaton potential $\psi$ is given by
\eqn\psifs{ \psi = -\biggl({\partial F\over \partial D}\biggr)_{T_W}
=\lambda} and the entropy is
\eqn\Sfs{S = -\biggl({\partial F\over \partial T_W}\biggr)_D= 0}
Flat space therefore has zero entropy as one might have expected.
However, it does not have zero energy, since
\eqn\Energy{ E = F+ST_W}
it follows that
\eqn\Efs{ E_{fs} = -4\lambda e^{\phi_W}}
This energy does not gravitate. At first sight this seems a bit paradoxical,
however the presence of a long-range scalar field causes the
weak equivalence principle to be violated. This energy is an
irreducible energy
which is tied up with the existence of the background
dilaton charge.

One might be concerned that our definition of the dilaton current was
quite arbitrary since we could replace the $e^\phi$ term in \dilcurr~
by any function of $\phi$ and still find a conserved current. If we did
this the new dilaton charge would just be a function of the old
dilaton charge and so there would not be any new conserved quantum
numbers associated with it. It should also be noted that this charge
is quasi-topological in that it arises solely from a boundary contribution.

Now we turn our attention to the Euclidean black hole solution \eubh~
Suppose that the wall is a distance $r$ from the event horizon.
The periodicity of the $\tau$-coordinate at $r$ is as we saw $T_c^{-1}$.
However, the periodicity of the wall of the box there is $T_W^{-1}$,
and so $T_W$ is related to $T_c$ by  Tolman relationship
\ref\Tolman{ R.C. Tolman, {\it Relativity, Thermodynamics and
Cosmology}, Oxford University Press, Oxford, 1934.}
 \eqn\tolman{ T_W = T_c\coth(2\pi r T_c). } As in the flat space case,
 we can straightforwardly evaluate the action and find in terms
of the coordinates that
\eqn\actionbhsss{I = 4\pi e^{\phi_0} \bigl(1 - 2\cosh^2(2\pi r T_c)\bigr).}
The dilaton charge in the box is
\eqn\Dbh{ D = e^{\phi_0} \cosh^2(2\pi r T_c).}
Consequently, the Helmholtz free energy is, when expressed in terms of the
canonical physical variables, $D$ and $T_W$,
\eqn\Fbh{ F=-4\pi D\biggl(T_W + {T_c^2\over T_W}\biggr).}
The chemical potential associated with the  dilaton charge is
\eqn\psibh{ \psi = 4\pi T_W\biggl(1 + {T_c^2\over T_W^2}\biggr).}
The entropy $S$ of the spacetime inside the box is then easily found
and is given by \eqn\Sbh{ S=4\pi D \biggl(1 - {T_c^2\over T_W^2}\biggr).}
Since $T_W > T_c$, this entropy is positive as one should expect.
The energy can now be found from \Energy~ and is
\eqn\Ebh{ E_{bh} = -8\pi D {T_c^2 \over T_W}. }
This represents the combined energy of the background field and
the black hole. If we re-write this using the same variables as the
flat space example, we find that
\eqn\ebh{ E_{bh} = -4\lambda e^{\phi_W} {T_c\over T_W}}
since  $\lambda = 2\pi T_c$.
It is now easy to extract that part of $E_{bh}$ that is the energy
of the black hole, $M.$ Recalling \tolman~ we find that
\eqn\enrgy{M= E_{bh} - E_{fs} = 4\lambda e^{\phi_W}(1-{T_c\over T_W})}
This quantity we interpret as being the energy of the black hole
itself, or equivalently its rest mass.
Furthermore, as we might have expected, in the asymptotic region
where $r\rightarrow\infty $, this expression reduces
 to
\eqn\madm{M=2\lambda e^{\phi_0}}
which is the ADM mass
of the black hole,
as described in reference \Wittenbh. The easiest way to see this is to
adapt the derivation of Arnowitt, Deser and Misner,
 \ref\adm{R. Arnowitt, S. Deser and C.W. Misner,{\it
\lq\lq The Dynamics of General Relativity"} in {\it\lq\lq Gravitation:
An Introduction to Current Research,} ed L. Witten, Wiley, New York,
1962.}.
The new feature of the analysis is the
introduction of the dilaton field. If we look at the volume
part of the gravitational action, \tgtspaction, the ADM mass is that part
of the corresponding  canonical Hamiltonian that is a boundary term. After
a short calculation, we discover that for a spacetime of the form
\genmetric, the ADM mass is given by
\eqn\admmass{ M_{ADM}= 2\Biggl(e^\phi f{\partial f\over\partial
r}\Biggr)_{r=\infty}} Thus for the black hole solution, we get
\eqn\admm{ M_{ADM}= 2e^{\phi_0} \lambda}
The ADM mass therefore agrees with the thermodynamic evaluation of the
energy.

Another way to write the mass of the hole
is in terms of the dilaton charge $D$ and the temperature $T_W$. In
this case, we find
\eqn\massss{M=8\pi DT_c\biggl(1-{T_c\over T_W}\biggr).}
This expression makes it clear that the range of $M$ is from
zero where $T_c$ tends to $T_W$, or alternatively when the black hole
is infinitely far from the wall, upto $8\pi DT_c$ where the wall of the
box tends to the event horizon.

Given \Sbh, one might think think that the entropy of the black hole
vanishes as one becomes infinitely far from it, that is
 where $T_W\rightarrow
T_c$. This would be in contrast to the four-dimensional case where
if one is infinitely far from the black hole, the entropy is given
precisely by \SBek. However, this is rather misleading as the entropy
should be expressed as a function of the energy of the black hole
and its dilaton charge as these specify the physical state of the black
hole. Writing $S$ as a function of $M, D$ and $T_c$, we find
\eqn\Smd{S={M\over T_c}\Biggl(1-{M\over16\pi DT_c}\Biggr)}
Since $M$ lies in the range $0\leq M\le 8\pi DT_c$,
the entropy of the hole for fixed $D$ and $T_c$ reaches a maximum
of $M/2T_c$ when $M=8\pi DT_c$, and has a minimum of zero when the
mass of the black hole vanishes.

Finally, we would like to find a formula analogous to \SBek. Such a
formula would be valid at $r=\infty$, and would specify the entropy of
the black hole in terms of the mass. The easiest way to derive such a
formula us to start from \Sbh, and substitute \Dbh~ for $D$, and Tolman
for $T_W$. The result is then independent of $r$. Using \admm, a result
valid for the mass at spatial infinity only, we find that
\eqn\Sme{ S={M\over T_c}}
It is this final formula, \Sme~
that should be regarded as being the two-dimensional analog of \SBek.

\newsec{Electrification}
A more realistic fundamental theory  will include fermions and gauge
fields. In this section we examine some black hole solutions for
the heterotic string. These solutions seem to have been first discussed
by Nappi, Yost and McGuigan \ref\hethole{C. Nappi, M.D. McGuigan and S.
Yost, Institute for Advanced Study preprint, IASSNS-HEP-91/57.}.
In the bosonic sector,
the lowest order string beta-functions are very similar
to those for the closed bosonic string. However there are
Yang-Mills fields $A$
and corresponding field strengths $F$ belonging to some anomaly-free
gauge group $G$, which replace
the tachyon field. Unlike the case in ten dimensions, a complete list
of such groups is not known, but an anomaly free $G$ is $SO(24)$.
The beta-functions are, to lowest order
 in $\alpha^\prime$,\eqn\hbetag{ R_{ab} = \nabla_a\nabla_b\phi +
\half \Tr F_{ac}F_b{}^c }
\eqn\hbetaphi{R - (\nabla\phi)^2 - 2\bxy\phi - {1\over 4}\Tr F_{ab}F^{ab}
=c^\ast}
\eqn\hbetaA{\nabla_bF^{ab}+(\nabla_b\phi)F^{ab} = 0}
where the central charge is now
\eqn\hcent{ c^\ast = {2c+3d-30\over 6}}
These equations can be derived from the target space action
\eqn\hact{I=\int_{\cal M}d^2x g^{1\over 2}e^\phi(c^\ast - R- (\nabla\phi)^2
+{1\over4}\Tr F^2) - 2\int_{\partial\cal M}e^\phi K~d\Sigma}
Evaluating this action for a field obeying the beta-function equations
yields an on-shell action that is again a pure boundary term,
\eqn\honsh{I=-2\int_{\partial\cal M}e^\phi(K+n^a\nabla_a\phi)d\Sigma}
with $n^a$ being the unit normal to the boundary $\partial\cal M$.

It is now a straightforward task to solve these equations
and find a black hole solution in exactly the same way as we did
for the bosonic string. We will take the gauge field to lie in a
$U(1)$ subgroup of the full gauge group. The solutions
that emerge are in some sense analogs of the Reissner-Nordstrom solution
in general relativity.
The spacetime metric is
\eqn\hbhole{ ds^2 = -{ (m^2-q^2)\sinh^22\lambda r\over\bigl(m +
(m^2-q^2)^\half \cosh2\lambda r\bigr)^2}dt^2 + dr^2}
with the dilaton field given by
\eqn\hdil{\phi=\phi_0+\ln\biggl(\half ({m\over
(m^2-q^2)^\half}+\cosh2\lambda r)\biggr)} and the vector potential one-form
by \eqn\ha{A={\sqrt2q\over m+(m^2-q^2)^\half\cosh2\lambda r}dt}
This solution appears to depend on four arbitrary parameters $\phi_0,q,m$
and $\lambda$. It should be noted that we have chosen our parameters
here in such a way that if one sets $q=0$, one obtains the previous
expressions given for uncharged black holes.
The parameter $q$ is related to the charge contained in the spacetime,
and is required to obey $q^2 < m^2$ so that we have a black hole
in an asymptotically flat spacetime,
rather than a naked singularity. It should be noted that the case where
$q=\pm m$ looks degenerate in these coordinates, but this degeneracy can
be removed by a suitable redefiniton of the $t$ coordinate. However, if
we do this, the spacetime fails to be asymptotically flat.

The charge Q inside the a box whose wall is located at $r$ is given
by the value of $-e^\phi\ast F$. Here
the star is the Hodge dual operator acting on the field strength 2-form
$F$. In terms of components, we could write
this as
\eqn\haltQ{Q = \half e^\phi \epsilon_{ab}F^{ab}}
It follows from
the analog of Maxwells equation \hbetaA, that this is a conserved charge.
Evaluating $Q$ for the
solution given here, we find that it is
\eqn\hQeval{Q = \sqrt2\lambda q
e^{\phi_0}{\sinh2\lambda r\over m+(m^2-q^2)^{\half}\cosh2\lambda r}.}
$\phi_0$ is an arbitrary constant related to the ADM mass of the hole
which can be calculated by the same method as given in section three.
It is
given by \eqn\eadmm{M_{ADM}={2\lambda\over(1-{m^2\over q^2})^{1/2}}
e^{\phi_0}.} $\lambda$ is
fixed by the central charge since
 \hbetaphi~ requires that
\eqn\hcent{c^\ast = -4\lambda^2}

This solution can be Euclideanized by the substitution
$t \rightarrow i\tau$ and $q\rightarrow ik$. In addition to the time coordinate
$t$, the parameter $q$ must be
Wick rotated because it appears as the time component of a vector.
Thus the Euclideanized version of the solution is
\eqn\hEbhole{ds^2 = {(m^2+k^2)\sinh^22\lambda r \over
\bigl(m+(m^2+k^2)^\half\cosh2\lambda r\bigr)^2}d\tau^2 + dr^2}
with the dilaton being given by
\eqn\hEdil{\phi=\phi_0 +\ln\biggl(\half({m\over(m^2+k^2)^\half}
+\cosh2\lambda r)\biggr)}
and the vector potential by
\eqn\hEa{A={\sqrt2k\over m+(m^2+k^2)^\half\cosh2\lambda r}d\tau.}
There is a conical singularity at $r=0$ unless $\tau$ is identified
with period $\beta$, where
\eqn\hHawT{\beta={\pi\bigl(m+(m^2+k^2)^\half\bigr)\over\lambda(m^2+k^2)^\half}}
To find the Hawking temperature, we take the inverse periodicity
and analytically continue back to the physical region by substituting
$k\rightarrow-iq$. Thus the temperature of the black hole is
\eqn\hTc{T_c={\lambda(m^2-q^2)^\half\over\pi(m+(m^2-q^2)^\half)}
}It should be noted that the temperature is zero in the case $q=\pm m$.
This extreme black hole is rather like the extreme Reissner-Nordstrom
solution, and has similar global structure. By analogy with the
uncharged case, there is a dilaton charge given by
\eqn\hDil{D=\int \epsilon_{ab} \nabla^be^\phi d\Sigma^a}
Evaluating this for the solution in question yields
\eqn\hDileval{D=\half
e^{\phi_0}\biggl ({m\over(m^2-q^2)^\half}+\cosh2\lambda r\biggr)} for the
dilaton charge enclosed by a box at a value of the radial coordinate given
by $r$.

We can now evaluate the  action as we did for
the uncharged black hole discussed in section three. The
Euclidean action $I$ is
\eqn\HEuclI{I=-2\pi\cosh(2\lambda r)
e^{\phi_0}\biggl({m\over(m^2+k^2)^\half}+1\biggr)} To compute thermodynamic
quantities, we must use this result to evaluate the free energy F in terms
of  the physical variables $T_W,Q,D$ and $\lambda$ after
continuing back to the physical region. It should be noted that $\lambda$
is fixed by the central charge. The temperature
of the wall of  the box is redshifted by the Tolman factor, so that we
find
$T_W$ is related to $T_c$ by
\eqn\hTwall{T_W = T_c\Biggl({m+(m^2-q^2)^\half\cosh2\lambda r
\over(m^2-q^2)^\half\sinh2\lambda r}\Biggr).}
Then from \hTwall, \HEuclI, \hDileval~ and \hQeval~
that the best way to construct the free energy $F=F(\lambda,D,T_W,Q)$
is parametrically by writing $F=F(\lambda,D,T_W,x)$ and
$Q=Q(\lambda,D,T_W,x)$ with the parameter $x=\lambda r$. We thus find
\eqn\hfree{F = -4\lambda D\coth2x}
and
\eqn\Qdef{{Q\over D}={\lambda\over{\sqrt{2}}\pi^2}\biggl({\pi\sinh2x\over
T_W}-{\lambda\over T_W^2}\biggr)\biggl(\lambda^2-2\lambda\pi T_W\tanh
x\biggr)^\half.}
 To find the various thermodynamic potentials is now
straightforward. The electrostatic potential is
\eqn\electropot{\Phi=-\biggl({\partial F\over\partial
x}\biggr)_{\lambda,D,T_W}\biggl({\partial Q\over\partial
x}\biggr)^{-1}_{\lambda,D,T_W}} and  the dilaton potential is
\eqn\dilpot{\psi=-\biggl({\partial F\over\partial
D}\biggr)_{\lambda,T_W,x} - \biggl({\partial F\over\partial
x}\biggr)_{\lambda,D,T_W}\biggl({\partial D\over\partial
x}\biggr)_{\lambda,Q,T_W}.}
The entropy is given by
\eqn\hetS{S=-\biggl({\partial F\over\partial x}\biggr)_{\lambda,D,Q}
\biggl({\partial T_W\over\partial x}\biggr)^{-1}_{\lambda,D,Q}.}
Thus an explicit formula for $S=S(\lambda,D,T_W,x)$ can be found.
We will not record these rather complicated
and unilluminating expressions here.

It is however quite straightforward to evaluate the entropy as seen from
infinity in a way that is analogous to \Sme. Using the result for the
free energy, \hfree, the ADM mass \eadmm, we discover afer some
manipulation, that entropy as seen from infinity is given by
\eqn\Smee{ S= {{\pi M \over \lambda}(1+ (1-{q^2\over m^2})^{1/2})}}
This result is very similar to that for the Reissner-Nordstrom solutions,
the entropy is largest for electrically neutral black holes, and
descreases down to half that value as one charges up the black hole
to its maximal value of the charge where $\vert q\vert$ approaches $m$.

 We will however note that the limit of $m \rightarrow \pm q$
corresponds to infinite dilaton and electric charges, and correpsonds to
zero temperature. However, the limit $m=\pm q$ is unattainable since that
particular spacetime fails to be asymptotically flat.
This is because the infinite amount of energy involved
has curled up the spacetime at infinity. It is interesting to view this
as an explicit realization of the third law of black hole thermodynamics.
Perhaps the
easiest way to see this is to examine \hbhole~ and rescale the $t$
coordinate by a factor of $(1-{q^2\over m^2})^\half$. The new metric
is then
\eqn\limit{ds^2 = -\sinh^22\lambda r~dt^2 + dr^2}
Thus as $r\rightarrow\infty$ the metric fails to be flat. The
point $r=0$ will be a conical singularity unless $t$ is identified with
the appropriate period. We therefore conclude that the zero temperature
limit is not physical in this case. There are therefore no zero
temeperature solitons in this theory.

\newsec{Exact Solutions}

The spacetimes presented  so far can be regarded as approximations valid
in the $\alpha^\prime$ limit to some exact solutions to the beta-function
equations. If one wants to find the metric and dilaton fields to higher
orders in $\alpha^\prime$, one way to do this is to determine the
beta-function equations to as high an order in $\alpha^\prime$ as one
can, and then solve the resulting field equations. By following such a
prescription, Jack Jones and Panvel \ref\JJP{I. Jack, D.R.T. Jones and J.
Panvel, University of Liverpool preprint, LTH-277, 1992} have shown that
the solutions presented in section two are exact for the type II
superstring. We therefore confident that our results hold for realistic
string theories. However, for the bosonic and heterotic strings, there
will be higher order corrections to the solutions outlined above. For the
bosonic string, we can still perform the exact calculation by following a
trick of Witten \Wittenbh. Suppose one starts from the WZW model at level
$k$ for $g ~\epsilon~ SU(1,1)$, we have then a conformal field theory with
a three dimensional target space. the action is
\eqn\wzwact{I= \half\int {\rm Tr} g^{-1}dg \wedge g^{-1}dg +
{ik\over 24\pi}\int {\rm Tr}(g^{-1}dg)^3}
Suppse that one now gauges an antidiagonal $U(1)$ subgroup $h$ by
\eqn\antidiag{g \rightarrow hgh}
The result is a confomal field theory with a two-dimensional target space
with a metric of the form \genmetric where now
\eqn\fofr{ f(r) = \biggl( \coth^2\lambda r - {2/k}\biggr)^{-\half}}
and
\eqn\phiofr{ \phi = \phi_0 + \ln{\sinh\lambda r\over f(r)}}
\ref\vvd{R. Dijkgraaf, E. Verlinde and H. Verlinde, Institute for
Advanced Study preprint, IASSNS-HEP 91/22}. However, the basic change
is only in the details of the metric and not its overall character. Thus,
for mass scales large compared to $(\alpha^\prime)^{-\half}$, we expect
the results we have obtained to be quite valid. The metric and dilaton
specified by \fofr and \phiofr are conformally invariant, but with a
central charge specified by $k$. The central charge is \GKO given by
\eqn\exactc{ c = {3k\over k+2 } - 1}
thus we will have a critical bosonic string theory if $k={9/4}$.
If we keep away from this value, there will be black hole solutions.
If $k={9/4}$, then $\lambda=0$ and the only metric will that of flat
space.
It would nice to be able to carry out our thermodynamic calculations in
the exact theory. However, we have no way of finding the spacetime action
for the exact theory, and so we cannot find the free energy, and hence we
cannot find the entropy. It would be intriguing to calculate these
quantities for an exact theory.

Our results show a remarkable parallel with similar results in four
dimensions even though the presence of the dilaton field causes
considerable complications. This makes it quite plausible to suppose
that these two dimensional stringy black models really do have sufficient
in common with realsitic examples to give some guide as to how to resolve
the problems of quantum gravity by extrapolation from these highly
simplified toy models.

\newsec{Acknowledgements}
MJP would like to thank Prof. I.M. Singer and the Mathematics department
at MIT for their hospitality during the completion of this work, which
was supported in part by US DOE grant DE-FG02-88ER25066.

\listrefs
\bye